\documentclass{article}

\newcommand{\pion}{\ensuremath{{\pi^0}}}

\newcommand*{\meson}[1]{\ensuremath{#1^\pm}}
\newcommand*{\quarkpair}[1]{\ensuremath{#1\bar #1}}
\newcommand*{\quark}[1]{\ensuremath{#1}}
\newcommand*{\Quark}[1]{\ensuremath{#1}-quark}
\newcommand*{\GeV}[1]{\ensuremath{#1\hbox{\hskip5pt GeV}}}
\newcommand*{\MeV}[1]{\ensuremath{#1\hbox{\hskip5pt MeV}}}
\newcommand*{\Approx}[1]{\ensuremath{\approx #1}}
\newcommand*{\ApproxP}[1]{\ensuremath{\approx #1 \%}}

\begin{document}

\begin{center}
{\Large Harmonic quarks and their precise masses}\par\bigskip
{\large Oleg~A.~Teplov}\par\smallskip
Institute of Metallurgy and Material Science 
of the Russian Academy of Science, Leninski prospekt 49, Moscow, 119991, Russia.\par
e-mail: teplov@ultra.imet.ac.ru
\end{center}

\vspace{4ex}
\centerline{\large\bf Abstract}
\medskip

An examination of charged two-quark meson masses hinted that the mass ratio
of neighboring quarks could simply be a constant. The concept of a 
harmonic quark oscillator based on a quark-antiquark pair is introduced. 
Unstable symmetric state of the harmonic quark oscillator can be broken 
to an asymmetric state with the formation the neighboring quark of lesser 
mass due to a weak reaction. A new recurrent equation of quark masses 
is obtained and the model of harmonic quark family is developed. 
The quark masses in the model are bound together into one rigid chain. 
With the electromagnetic contribution taken into account the quark masses
are calculated with the 0.03 percent inaccuracy. It follows from the model 
that the muon is a single u-quark mass state fixed as a lepton. The neutral 
pion is probably a stable harmonic oscillator based on a quark-antiquark 
u-pair.

\bigskip

%% --------------------------------------------------

\section{Introduction}

Let us briefly review the present state of the quark mass problem,
considering that, modern physics has enough experimental data to be 
certain in existence of quarks. Quarks acquire their masses in a process
which is related to the Higgs sector and goes back to the original Higgs's
work of 1964~\cite{higgs}. Quarks are interacting with the Higgs boson 
(bosons) proportionally to masses of quarks. The extensive review of 
the Higgs boson search and theories presented in~\cite{carena}. 
The Higgs boson is believed to have the mass \GeV{115}~\cite{aleph} or 
perhaps even up to \GeV{200}~\cite{aleph-delphi}. In view of indeterminacy 
of a theoretical position, both the number of bosons and their masses remain 
unknown and therefore the concrete mechanism of quark mass formation is also
unclear. As twenty years ago ~\cite{okun} and now ~\cite{peskin} practically 
the same opinion is stated: \par``... we do not at all understand the lepton 
and quark mass spectrum ...'' ~\cite{okun}\par
``The fact, that the mass spectrum of quarks and leptons envelops 5
orders of magnitude, is a mystery from any point of view... Even with
universal recognition of standard model, these problems seem very far
from resolution'' ~\cite{peskin}. Theory urges us to employ the concept 
of a quark mass with a great deal of discretion, because only bound 
quark systems (i.~e.~particles) are observable. Depending on the 
concrete character of quark interaction, we can expect either positive 
or negative binding energy. The negative binding energy is analogous to 
mass defect in nucleus when the mass sum of the component quarks is 
more than the mass of the relevant hadron. The positive binding energy 
signifies that in addition to quark masses there is an additional energy
in a hadron, for example, the energy of color field, which provides the 
quark confinement. The notion of quark mass and its value depends upon the 
theoretical model used in a given experimental or theoretical work.  
With respect to quark mass modern physics developed a number of notions: 
there exist bare quark mass, physical quark mass, constituent quark mass 
and effective quark mass. Effective quark mass is inherently related to 
the impulse used to calculate it~\cite{peskin} and we won't consider it 
here.
An interacting field theory has a well-defined notion of the rest mass of
a single particle state, which is called physical mass.  There can
also be discrete bound states of the particle-antiparticle pairs.

The mass parameter in the QCD Lagrangian is considered to be a bare quark
mass~\cite{peskin}.  The various estimations within the QCD theory using the
$\overline{MS}$-schemes and lattice simulations show that the bare masses of
the light \quark{u}- and \quark{d}-quarks are less than \MeV{10}~\cite{PDG}.
The estimation of the bare mass ratios for the \quark{u}-, \quark{d}- and
\quark{s}-quarks is $1:2:40$ respectively. Therefore, it is accepted, that
the fundamental Lagrangian of strong interaction does not contain any hint
on a symmetry with respect to quark masses ~\cite{peskin}.

Approximate constituent quark masses can be obtained from experimental data of
the hadron masses~\cite{bopp}.  Constituent quark is deemed to be formed
from the bare quark and its ``coat'', i.e.~surrounding gluon field
energy, the sea quarks and all other unsolved hadron structure problems.
First of all, the ``coat'' notion makes the concept of a constituent
quark mass fuzzy, because the quark's incapability of solitary existence 
makes the mass of it's ``coat'' dependent on a particular environment, 
i.e.~ from the quark structure of hadron and the quark-gluon dynamics.

On the other hand, the mass discreteness of the light hadrons with equal
quantum numbers is several times greater than pion masses. It is difficult
to understand why a light constituent quarks of pions have to dress in 
additional heavy ``coat''. At the same time significant discreteness of the
hadron mass spectrum makes it possible to hope that the quark-gluon field 
energy is also subject to quantization.

On the whole, after the forty years the birth of the quark model, the 
solution of the quark mass problem seems to be far still. It is also
apparent that both the hadron internal energetic structure and quark-quark
interaction dynamics is impossible to fully understand without the knowledge
of quark mass quantization and the precise values of their spectrum. 
These reasons and the author's conviction in the existence of the 
analytic relation for quark masses gave rise to the present work. 
Hereinafter under a quark mass we shall understand a physical rest mass of 
the single particle state of some interacting quantum field.

\section{Quark properties and experimental data}
As a first stage of the study we shall ascertain the fact that there is
both theoretical reason and experimental data in favor of certain
relation between quark masses. However, let's begin with a few words about
flavor as it is the only known quantum number, which is obviously
connected with quark mass.  All other quantum numbers is strictly
quantized and are either constant or periodic by quark generations.
Let's ask ourselves, is quark mass quantized?  It evidently is, for all
available energetic information about matter indicates that quantum
effects and energy discrete levels became dominant as an object
dimensions decrease.  The mass spectrum of the hadrons with different
flavors support the idea of quark mass quantization.  Therefore another
question looms: is quark mass strictly quantized? According to
field theory, the answer is yes.  The flavor quantum number is
essentially a reflection of quark's internal energy---its physical mass.
Let's see how the flavor changes in the basic weak interactions of quarks.
\begin{eqnarray}
  Q_n + W^\pm &\leftrightarrow& Q_{n+1}    \\
  Q_n + e^\pm &\leftrightarrow& Q_{n+1} + \nu
\end{eqnarray}

These are transition reactions between quarks with neighboring
flavors. It is irrelevant whether real or virtual bosons the quarks are
interacting with.  The point is that both the quark's mass and charge
are changed.  Therefore, two successive transformations will result in
double mass transformation with the charge remaining constant---that is,
flavor will be the only quantum number that will change.

We shall now examine theoretical computations of bare quark masses 
and certain mesons ~\cite{PDG}.  The values of bare 
quark masses are known with one digit precision at best. Let's select 
for the analysis the charged mesons consisting of two neighboring 
quarks $Q_n$ and $Q_{n+1}$ with minimal mass in their ground state.
In the table~1 these masses are given together with their relations
\footnote{The table~1 data are taken from the Particle Data Group listing 
~\cite{PDG}}.

\par\bigskip

\begin{center}
  Table 1.  Masses of bare quarks and certain mesons.
  \medskip\small

  \begin{tabular}{|c|c|l|c|c|c|}
    \hline
    quark & quark mass & quark mass & meson        & meson mass & meson mass \\
    & MeV        &\hfil ratio\hfil&   & MeV        &  ratio\\
    \hline
    \quark{d}  & 3--9 &\hfil-\hfil& - & - & - \\
    \quark{u}  & 1--5 &\Approx{0.5}\hfill(\quark{u/d})&
    \meson{\pi}(\quark{du})& 139.57 & - \\
    \quark{s}  & 75--170 &\Approx{40}\hfill(\quark{s/u})&
    \meson{K}(\quark{us})& 493.64 & 3.54 (\meson{K}/\meson{\pi})\\
    \quark{c}  & 1150--1350 &\Approx{10}\hfill(\quark{c/s})&
    \meson{D_s}(\quark{sc})& 1968.5 & 3.99 (\meson{D_s}/\meson{K})\\
    \quark{b}  & 4000--4400 &\Approx{3.4}\hfill(\quark{b/c})&
    \meson{B_c}(\quark{cb})& 6400 & 3.25 (\meson{B_c}/\meson{D_s}) \\
    \quark{t} & 174300 &\Approx{42}\hfill(\quark{t/b})& - & - & - \\
    \hline
  \end{tabular}
\end{center}

\bigskip

The bare quark mass ratios do not show any regularity but the
meson mass ratios definitely present certain degree of uniformity.  The
mean value of their mass ratios is 3.6.  Assuming that the neighboring
quark mass ratio is actually a constant of 3.6 we can obtain the
following quark masses ({\em starting from the mean table value for the
\Quark{b}}): 4200(\quark{b}), 1167(\quark{c}), 324(\quark{s}), 90(\quark{u}) 
and 25(\quark{d}).  The sum of the neighboring quark masses will be
about 17\% less than the meson masses for all considered mesons
except \meson{D_s} (\ApproxP{25}).  This sounds promising, for after a
proportional increase of the quark masses we'll have a match for three
of the four examined mesons.  Thus, taking the \Quark{b} mass value
equal to \MeV{5000} we'll have for the mesons \meson{\pi}, \meson{K} and
\meson{B_c} the mass values of 137, 493 and \MeV{6390} respectively.

Other examples that meson mass ratio is roughly constant are given in
the table 2.

\begin{center}
  Table 2.  Meson masses and mass ratios
  \medskip
  
  \begin{tabular}{|c|c|c|}
    \hline
    meson & meson mass, MeV & meson mass ratio\\
    \hline
    $\omega$ & 782 & - \\
    $J$ &     3097 & 3.96 $(J/\omega)$\\	 
    $\mathit\Upsilon(1S)$ & 9460 & 3.05 $(\mathit\Upsilon/J)$\\
    \hline
    $\varphi$ & 1019.4 & -\\
    $\mathit\Psi(2S)$ & 3686 & 3.61 $(\mathit\Psi/\varphi)$\\
    $\mathit\Upsilon(2S)$ & 10023 & 2.72 $(\mathit\Upsilon(2S)/\mathit\Psi)$\\
    \hline
  \end{tabular}
\end{center}

In contrast to table~1 the mesons given here have a hidden flavor (that
is, a quark-antiquark pair).  Thus, we can claim that this experimental
data set also supports the idea of approximate constancy of the neighboring
quarks mass ratio (\Approx{3.6}).

So then, there is certain multiplicative pattern in mass transformation
between quark flavors.  This probably means that quarks are in a
similar state in these mesons with various masses.  This is quite hard
to explain considering that quark and meson masses differ by two orders
of magnitude.  The light quarks of a pion should be relativistic
whereas the heavy quarks of a heavy meson are not very relativistic even
when taken at their bare masses.  There would be no such contradiction if
the quark masses themselves have the scale similarity with the same
factor that charged mesons do. Then the quarks should not be very
relativistic in all these mesons.

\section{Construction of the quark mass model}

Now we will attempt to build a quark transformation model and to 
calculate the quark mass transformation factor.  We should investigate 
the quark interaction process that is essentially to guess the new key 
moments of the strong interaction.  A couple of simple question to start 
things off: first of all---what is the character of quark-antiquark 
interaction, and second of all---what is the result of that interaction?  
It is rather easy to answer the second question.

This interaction produces either a meson based on the quark-antiquark
pair ({\em e.g. a vector meson}) or the complete annihilation of the
pair with the birth of photons or smaller mass quarks and other particles.
%%%%%%%%
The formation of vector meson has only deferred the quark-antiquark 
annihilation. There are seemingly two extremes---either the pair is  
present ({\em meson}) or it disappears ({\em annihilation}). But this 
``yes or no'' situation can be avoided by introducing the concept 
of a discrete annihilation spectrum or discrete coupling of a 
quark-antiquark pair.
It is not much of a novelty actually.  A good old discrete bound states
of an interacting field theory, which are lower than the doubled
physical mass of the quark.  Bound quark-antiquark states are well known
(e.g., their paired states are presented in table~2).  We are approaching 
the main question now---what is the binding energy value of quark-antiquark 
bound state and what quantum laws operate by the levels of binding energy?
It is still an open question now as we poorly know even the quark masses. 
It is similar a movement on the closed circle and our chance only to guess.
Then new variant of the answer is below stated.
\par\medskip {\bf Proposition \#1}
\par {\em The quark-antiquark pair of the same flavor ~$n$ can, instead 
of the complete annihilation, annihilate partly to the bound state of a
\textbf{harmonic oscillator} with the mass equal to $m_n\cdot{4\over\pi}$.}

The partial annihilation means that the quark pair with total mass $2m_n$ are
involved in such an interaction with the following equation for their
bound state with total energy $m_{sum}$:

\begin{equation} \label{HO}
  m_{sum}=2m_n\int_0^\pi {sin(x)\over\pi}dx={4\over\pi}m_n
\end{equation}

This process (let us name it the {\em harmonic quark oscillation}) can
be also interpreted as a harmonic oscillation of the quark-antiquark 
pair over space-time from complete separation of quarks to their complete 
annihilation. So for example, the time-dependent quadratic mass term can 
be noted as the equation:    

\begin{equation} \label{mtau}
  m^2(\tau)=(2m_n)^2cos^2(\omega\tau)
\end{equation}

The time-averaged mass will also be equal to $m_n\cdot{4\over\pi}$.
In eq.~(\ref{mtau}) we are not considering the transition of the mass into
other energy forms and these oscillations should be forbidden by the 
law of energy conservation. Nonetheless in the virtual meaning with the 
help of uncertainty relation we have a right to consider them to a 
certain boundary frequency $\omega$.

The equation~(\ref{HO}) is also corresponds to a stable state of the
quark pair with a certain mass defect value (\ApproxP{36}).  This state
could be similar to quarkonium solution of the wave equations with
quarks mutually revolving on some stable orbit, except that it has the very
high binding energy.

At this stage, it's important to examine the stability of the harmonic
oscillation.  The process defined by~(\ref{mtau}) would not be stable
for exactly the same reason which has allowed not to forbid it. Namely,
that we are already at the energy fluctuation level comparable to the quark
mass in the context of uncertainty relation, which means that the other
fluctuations would immediately destroy the harmonic oscillator.  But
what will follow the destruction?
%%% overcome
There are at least two ways to overcome this instability.  The first 
one is that an unstable annihilating oscillator is fast damped and then 
quark pair goes to mentioned stationary state, which corresponds to 
equation~(\ref{HO}). The second way is to accept the proposition two.

\par\medskip {\bf Proposition \#2}\par {\em The labile symmetric state
of the harmonic annihilating oscillator based on a quark-antiquark pair
of the same flavor can be broken in one or another way and would 
subsequently pass into the asymmetric state based on quarks of different 
kinds, i.e.\ consisting of a quark-antiquark pair with neighboring 
flavors.}\par\medskip 
To clarify this proposition let us write one of the possible reaction \\
of \quark{u}-antiquark birth.

\begin{equation} \label{ubirth}
  (\quarkpair{s})_{ho}+e^-=s+\bar u+\nu
\end{equation}

where $(\quarkpair{s})_{ho}$ is a harmonic oscillator based
on a strange quark-antiquark pair and $\nu$ is a neutrino.  Reactions
similar to~(\ref{ubirth}) can be written down for any quark harmonic
oscillator, both with actual and virtual electrons.  Doubling the
number of participants we can write the reactions with quarks only:

\begin{equation} \label{sspair_break}
  2(\quarkpair{s})_{ho}=s+\bar s+u+\bar u
\end{equation}

This reaction demonstrates the birth of the quark-antiquark
\quark{u}-pair from two harmonic oscillators based on strange
quark-antiquark pairs.  The main point here is the possibility of such
reactions with neither extra participants nor energy addition.
%%%
Note that, it is weak interaction with participation of other particles 
that is responsible for the symmetry breaking in one harmonic oscillator 
with the subsequent transition into the asymmetric state.

Now we can write down a recurrent relation between quarks of neighboring 
flavors, considering that a harmonic state with the energy value of
${4\over\pi}m_n$ generates two quarks with masses $m_n$ and $m_{n-1}$:

$$
m_{n-1} = {4\over\pi}m_n - m_n = m_n({4\over\pi}-1) = 0.27323954m_n
$$

\begin{equation}\label{MQ}
   {m_n\over m_{n-1}}= {\pi\over(4-\pi)} = 3.6597924 \stackrel{\rm def}{=} MQ
\end{equation}

Thus, having introduced two propositions we managed to obtain an
equation, which links together the masses of neighboring quarks.
Hereafter using the word quarks, we shall mean the harmonic quarks,
i.e.\ the quarks, which subject to the equation~\ref{MQ}.

Let's summarize the implications of the equation~(\ref{MQ}):

\begin{itemize}
\item Quarks form a rigid chain and their masses are all bound together
\item The quarks have a logarithmically equidistant mass spectrum
\item The multiplication factor $MQ$ between neighboring quark masses
    is equal to $3.66$ (\textit{which is practically equal to the mean
  meson mass ratio (3.6) mentioned above})
\item A quark flavor uniquely determines the transformation number
\end{itemize}

To stress this last point, the~(\ref{MQ}) can be written as
\begin{equation}\label{qmass}
  m_n = m_0\left[{\pi\over4-\pi}\right]^n
\end{equation}
where $m_0$ is a hypothetical initial quark mass, and the quark order
number $n$ is essentially its flavor.
Thus we have:
\begin{center}
  \begin{tabular}{r|ccccccc}
    n     & 1 & 2 & 3 & 4 & 5 & 6 & 7\\
    \hline
    quark & $d$ & $u$ & $s$ & $c$ & $b$ & $t$ & $b'$
  \end{tabular}
\end{center}

It's worthwhile to note that the total mass of any neighboring quark
pair is precisely equal to the mass of the harmonic oscillator based on
the quark with greater mass.  Moreover, with respect to a given quark
its two neighbors can be considered as an upward excitation (the quark
with greater mass) and a downward excitation (the quark with lesser
mass), with the equal electric charges of both excitations.

Finally, we have the chance to calculate actually the genuine quark
masses and afterwards to verify the validity of the original
propositions using ample experimental data.

\section{Harmonic quark masses}

\subsection{Quark masses estimation}
We can estimate the harmonic quark masses assuming that the charged
meson mass is a total mass of the valence quarks composing it.  Table~3
presents these estimations for each of the four stable mesons from the
table~1.

\begin{center}
  Table 3.  Harmonic quark masses, MeV (\textit{zero approximation})
  \par\medskip
  \begin{tabular}{|l|c|c|c|c|c|c|}
    \hline
    meson & $d$ & $u$ & $s$ & $c$ & $b$ & $t$\\
    \hline
    \meson{\pi}($ud$) & 29.9 & 109.6 &   401 & 1468 & 5373 & 19665\\
    \meson{K}($su$)   & 28.9 & 105.9 & 387.7 & 1419 & 5193 & 19005\\
    \meson{D_s}($cs$) & 31.5 & 115.4 & 422.4 & 1546 & 5658 & 20708\\
    \meson{B_c}($bc$) & 28.0 & 102.5 & 375.3 & 1373 & 5026 & 18396\\
    \hline
  \end{tabular}
\end{center}

The dispersion of these estimations does not exceed 12\%.  It is worth
mentioning that \meson{B_c} meson mass value is measured with 8\%
inaccuracy and the estimation for meson \meson{D_s} supposedly produce
an overestimation\footnote{The reasons for this overestimation are
beyond the scope of the present paper and will be explained in further
publications.}.  The pion- and kaon-estimated masses differ by 3.5\%.
This difference has a natural explanation if the sizes \meson{\pi} and 
\meson{K} are approximately equal. The only difference 
of their structure is in \quark{d} and \quark{s}-quarks and the localization 
of the \Quark{d}, which is different by an order of mass magnitude from the 
\Quark{s} will demand an additional kinetic energy. That's why the 
pion-estimated quark masses will be greater.

\subsection{Bounds of harmonic quark masses}

Keeping in mind that quark masses in the model are rigidly bound, the
heavier mesons we use for the quark mass determination, the more
accurate it will be.  Therefore we would start with the heavy charged 
two-quark meson with the mass measured exact enough, for example, \meson{b} 
with the mass \MeV{5279.0\pm0.5}.  The kaon-estimated (table 3) total 
mass of \quark{b} and \quark{u}-quarks is \MeV{5299}.  Provided that 
binding energy in the meson is not greater than the \Quark{u} mass, 
in other words not greater than 50\% of its mass per single quark 
(a binding energy in harmonic oscillator is only 36\% from mass of the 
everyone quark) we have that \meson{b} mass is an upper bound for 
the harmonic \Quark{b} mass.
The lower bound can be determined as follows.  Let's assume that the
value of binding energy in a meson is positive but less than a pion
producing threshold (about \MeV{140}).  Hence the lower bound of a
harmonic \Quark{b} mass is equal to $5279-140-109=\MeV{5030}$. Now we 
can calculate bounds for all other quarks using equation (\ref{qmass}). 
The calculated data are given in the table~4. 

\begin{center}
  Table 4.  Bounds and mean values of the harmonic quark masses, MeV
  \par\medskip
  \begin{tabular}{|r|cccccc|}
    \hline
    & $d$ & $u$ & $s$ & $c$ & $b$ & $t$\\
    \hline
    lower bound& 28.0 & 102.6 & 375 & 1374 & 5030 & 18409\\
    upper bound& 29.4 & 107.7 & 394 & 1442 & 5279 & 19320\\
    \hline
    \hline
    mean value& 28.7 & 105.2 & 384 & 1408 & 5154 & 18864\\
    \hline
  \end{tabular}
\end{center}
We may see that the bound values for quark masses situate inside same data
of tables 3 and the mean values are almost equal to kaon-calculated values.  
The fact that \meson{\pi} mass exceeds the sum of the upper mass bounds 
of the \quark{u} and \quark{d} by \Approx{\MeV{2.5}} means that pion 
binding energy is positive.  It means that quarks are completely
separated in \meson{\pi} and there is an additional energy of several 
MeVs.  What kind of energy it is?  The \quark{u} and \Quark{d}s of a 
pion charged alike and will repel each other.  The repulsion force will 
be the more effective the less distance is between the quarks.  On the 
other hand, the color field would act on the longer distances.  Thus 
the additional kinetic energy would be maximal at the color and Coulomb 
forces equality and would transform to Coulomb energy on the small 
distances and to color-field energy on large distances.  The value of 
this additional energy is naturally corresponds to the electromagnetic
split in hadronic duplets and triplets.  The complete separation of the 
quarks in \meson{\pi} and the structural similarity of the stable 
charged mesons discovered earlier makes it possible to suppose that the 
quarks in other charged mesons (in particular \meson{K} and \meson{b})
are also completely separated.

\subsection{Electromagnetic split is taken into account}
To calculate the mean value of electromagnetic split the author used
well-known mass differences of the charged and neutral particles up to
\MeV{1350}, i.e.\ of pions, kaons (with $J=0,1$) and basic baryon octet.
The seven differences altogether.  The mean split value is
\MeV{4.13\pm1.47}.  Decreasing by this the mass values of \meson{\pi},
\meson{K} and \meson{b} and using equation~\ref{MQ} we obtain the quark 
masses to a first approximation (table~5). The inaccuracy value given 
depends only on electromagnetic split measurement inaccuracy and the 
meson mass. (For the \meson{b} its mass measurement inaccuracy was also 
considered).

\begin{center}
  Table 5.  Harmonic quark masses (MeV) \par
  determined with electromagnetic split consideration\par
  (\textit{first approximation})
  \nopagebreak\par\medskip
  \begin{tabular}{|r|cccccc|l|}
    \hline
    meson & $d$ & $u$ & $s$ & $c$ & $b$ & $t$ & error,\%\\
    \hline
    \meson{\pi}&29.066&106.374&389.31&1424.8&5214.4&19083.7&$\pm$1.08\\
    \meson{K}  &28.706&105.058&384.49&1407.2&5149.9&18847.5&$\pm$0.30\\
    \meson{b}  &28.815&105.456&385.95&1412.5&5169.4&18919.0&$\pm$0.030\\
    \hline
  \end{tabular}
\end{center}

There are two positive moments to emphasize here.  
First of all, pion- and kaon-calculated quark masses differ only by 
1.2\%, which is three times less than the same difference of masses to a 
zero approximation and, moreover, is in accordance with the inaccuracy 
calculated.  
And second of all, the most accurate values of the harmonic quark 
masses, calculated with the \meson{b} meson (\ApproxP{0.03} inaccuracy), 
fall inside the range of pion- and kaon-calculated values.  Hereafter 
the phrase ``harmonic quark masses'' would relate to these most accurate 
values.

\section{First results}

We will now some demonstrate the ability of harmonic model to solve 
real-world issues. Even the most superficial examination of the harmonic quark
masses produce two impressive results:
\begin{itemize}
\item The total mass of \quark{u} and \quark{d} quarks is extremely
  close to the mass of neutral pion.
\item \Quark{u} mass is extraordinarily close to the muon mass.
\end{itemize}

Recollecting that neutral pion is the lightest existing hadron and
the truly neutral particle it seems that \textbf{neutral pion with
great share of probability is a stationary harmonic oscillator based on
quark-antiquark \textit{u}-pair}.  The pion mass (\MeV{134.97}) differs
from the oscillator mass (\MeV{134.27} by 0.5\% only.  It's likely that
the relativistic correction would account for that.

What is to \Quark{u} and muon, the difference between them is only
\MeV{0.2}, with the \Quark{u} mass value inaccuracy of \MeV{0.03} which
is seven times less than the difference.
The obvious conclusion is that \textbf{muon is a successful attempt of
Nature to explicitly fix the single \Quark{u} mass state as a lepton
suppressing color and fractional charge.}  As muon charge is greater by
$\frac{1}{3}e$ than the \Quark{u} charge, the Coulomb energy of muon should
also be greater. If electron mass is practically Coulomb energy then the 
charge in $\frac{1}{3}e$ add to the quark mass \Approx1/3 from the electron mass.  
Hence, the difference in \MeV{0.2} has an easy explanation. Thus the 
mystery of muon mass is naturally solved. It is strong result and good 
argument for harmonic model. 
Furthermore, the $MQ$ factor works well not only with quarks but with 
pions too, thereby supporting the idea of similar structural formation 
of the mesons with different flavors:
\begin{itemize}
\item $m_\pion\cdot MQ=\MeV{493.99}$, which is only \MeV{0.34} greater
  than the mass of the first meson containing strangeness (\meson{K})
\item $m_{\meson{\pi}}\cdot {MQ}^2=1869.4 {\rm(MeV)}$, which is precisely 
equal to the mass value of the first meson containing charm (\meson{D})
\end{itemize}
 In the last case, the additional energy of the pion quarks is also
transformed by the factor $MQ^2$, which leads to the effectively
increase of the \meson{D} meson mass.  This could be one of the
above-mentioned reasons of the harmonic quark masses overestimation when
calculated with \meson{D} and \meson{D_s} mesons.\linebreak

{\large \bf Acknowledgement} \\

I thank  ~V.~Teplov for the text correction and help with Latex.

%% -------------------------------------------------- 

\end{document}